\documentclass[sigconf]{acmart}
\usepackage{makecell}
\usepackage{multirow}
\usepackage{bbding}
\usepackage{xcolor,soul}
\usepackage{graphicx}
\usepackage{adjustbox}

\usepackage{array}
\newcolumntype{C}[1]{>{\centering\arraybackslash}p{#1}} 
\usepackage{tabularx} 
\usepackage{booktabs}

\usepackage{graphicx}  
\usepackage{caption}    
\usepackage{subcaption}  

\setcopyright{none}
\renewcommand\footnotetextcopyrightpermission[1]{}
\usepackage[utf8]{inputenc} 
\usepackage[T1]{fontenc}
\usepackage{booktabs}       
\usepackage{pifont}         
\newcommand{\cmark}{\ding{51}} 
\newcommand{\xmark}{\ding{55}} 
\usepackage{enumitem}

\AtBeginDocument{%
  }

\copyrightyear{2025}
\acmYear{2025}
\setcopyright{acmlicensed}\acmConference[MM '25]{Proceedings of the 33rd ACM International Conference on Multimedia}{October 27--31, 2025}{Dublin, Ireland}
\acmBooktitle{Proceedings of the 33rd ACM International Conference on Multimedia (MM '25), October 27--31, 2025, Dublin, Ireland}
\acmDOI{10.1145/3746027.3756869}
\acmISBN{979-8-4007-2035-2/2025/10}

\begin{document}

\title{AudioFab: Building A General and Intelligent Audio Factory through Tool Learning}

\author{Cheng Zhu}
\affiliation{
  \institution{Hunan University}
  \city{Changsha}
  \state{Hunan}
  \country{China}
}
\email{zhucheng@hnu.edu.cn}

\author{Jing Han}
\authornote{Jing Han and Zixing Zhang are corresponding authors.}
\affiliation{
  \institution{University of Cambridge}
  \city{Cambridge}
  \country{UK}
}
\email{jh2298@cam.ac.uk}

\author{Qianshuai Xue}
\affiliation{
  \institution{Hunan University}
  \city{Changsha}
  \state{Hunan}
  \country{China}
}
\email{xueqianshuai@hnu.edu.cn}

\author{Kehan Wang}
\affiliation{
  \institution{Hunan University}
  \city{Changsha}
  \state{Hunan}
  \country{China}
}
\email{wangkh@hnu.edu.cn}

\author{Huan Zhao}
\affiliation{
  \institution{Hunan University}
  \city{Changsha}
  \state{Hunan}
  \country{China}
}
\email{hzhao@hnu.edu.cn}

\author{Zixing Zhang}
\affiliation{
  \institution{Hunan University}
  \city{Changsha}
  \state{Hunan}
  \country{China}
}
\email{zixingzhang@hnu.edu.cn}
\authornotemark[1]

\renewcommand{\shortauthors}{Cheng Zhu et al.}

\begin{abstract}
Currently, artificial intelligence is profoundly transforming the audio domain; however, numerous advanced algorithms and tools remain fragmented, lacking a unified and efficient framework to unlock their full potential. Existing audio agent frameworks often suffer from complex environment configurations and inefficient tool collaboration. To address these limitations, we introduce \textit{AudioFab}, an open-source agent framework aimed at establishing an open and intelligent audio-processing ecosystem. Compared to existing solutions, AudioFab’s modular design resolves dependency conflicts, simplifying tool integration and extension. It also optimizes tool learning through intelligent selection and few-shot learning, improving efficiency and accuracy in complex audio tasks. Furthermore, AudioFab provides a user-friendly natural language interface tailored for non-expert users. As a foundational framework, AudioFab’s core contribution lies in offering \textit{a stable and extensible platform for future research and development in audio and multimodal AI}. The code is available at \url{https://github.com/SmileHnu/AudioFab}.
\end{abstract}




\maketitle

\section{Introduction}
\label{sec:intro}

In recent years, audio technology has become increasingly important in both daily life and industrial applications. Numerous powerful professional tools have emerged for tasks such as voice interaction, audio restoration, and music production~\cite{chae2023exploiting, alghamdi2022talking, ghosal2023text}. However, most audio tools are either single-function or combine multiple functions, making it hard to create a unified workflow that is both versatile and specialized. Complex, multi-stage tasks (e.\,g., end-to-end processing from audio analysis to music creation) often require switching tools and models based on different needs, complicating operations and reducing user-friendliness for non-professionals. Moreover, current intelligent solutions lack an efficient collaborative mechanism for task decomposition, tool selection, and execution feedback, hindering the integration of diverse audio tools into real-world workflows and failing to meet practical needs.

In this context, the advent of Large Language Models (LLMs) and the concept of \textit{Tool Learning}~\cite{toollearing} offers a new paradigm. By coordinating key stages of tool learning,
LLMs can use external tools to complete complex tasks, leading to the creation of LLM-based agents (e.g., METAGPT~\cite{METEGPT}). 

However, compared to rapid advancements in NLP~\cite{xi2025rise} and CV~\cite{wang2024genartist}, audio agent technology faces significant challenges.
First, the \textit{limited functional coverage} of current audio agents restricts their ability to meet the diverse demands of modern audio processing. Second, although many complex tasks can be solved with a few well-chosen tools, existing systems burden LLMs with entire tool libraries, resulting in \textit{excessive contextual information} and increasing the risk of interpretive errors (\textit{aka} tool hallucination), which may escalate into task hallucination and generate misleading outputs. Finally, tool modules in current systems are often tightly coupled and lack dedicated management mechanisms, making it \textit{difficult to add or update tools} without triggering dependency conflicts or version incompatibilities, thus hindering sustainable development.

To address these challenges, we present \textit{AudioFab, a general and intelligent audio factory designed for audio-centred multimodal applications}, which introduces improvements in \textit{functional breadth, tool management, and Tool Learning strategies.} 
The main contributions of this work are as follows:
\begin{itemize}[left=0pt, nosep]
    \item \textbf{A specialized audio tool library with modular management}: we collect a comprehensive set of audio-related tools for building a specialized audio tool library and introduce a modular management mechanism, supporting extensibility and mitigating dependency conflicts.
    \item \textbf{An intelligent tool learning process}: we develop a tool learning and selection strategy that dynamically adjusts context length based on user input and leverages few-shot learning to provide usage examples for selected tools, enabling LLMs to identify and invoke tools more effectively.
    \item \textbf{A unified and user-friendly audio agent framework}: we build an audio-specific agent framework using the Model Context Protocol (MCP), providing a unified interface and simplifying access for non-expert users via natural language interaction.
\end{itemize}

AudioFab is released as an \textit{open-source} toolkit under the \textit{CC BY-NC 4.0} license and actively maintained on GitHub. Built in Python, it integrates a range of dependencies, including PyTorch, audio processing libraries, and web frameworks. It supports \textit{easy installation} via a one-command Conda setup. The toolkit provides \textit{clear instructions} for adding new APIs or custom audio tools, and comes with \textit{comprehensive examples and documentation} in both English and Chinese, to support a broad user community.


\begin{table}[!t]
  \centering
\caption{\textbf{AudioFab and Comparable Software.} \cmark = supported, \xmark = unsupported, ``None'' = unknown, \textit{MM}: Module Management, \textit{Edt.}: Editing, \textit{Und.}: Understanding, \textit{Gen.}: Generation, \textit{Bat.}: Audio Batch Input, MCP: Model Context Protocol.}

  \label{tab:2}
    \begin{tabularx}{\columnwidth}{lcXXXXX}
      \toprule
       & \textbf{Platforms} & \textbf{MM}  & \textbf{Edt.} & \textbf{Und.} & \textbf{Gen.}  & \textbf{Bat.} \\
      \midrule
      AudioGPT~\cite{huang2024audiogpt}    & LangChain & \xmark & \xmark & \cmark & \cmark & \cmark\\
      WavCraft~\cite{liang2024wavcraft}  & None & \xmark & \cmark & \xmark & \cmark  & \xmark\\
      WavJourney~\cite{liu2025wavjourney}   & None & \xmark & \cmark & \xmark & \cmark & \xmark  \\
      \textbf{\textit{AudioFab (ours)}}  & MCPs & \cmark  & \cmark & \cmark & \cmark   & \cmark \\
      \bottomrule
    \end{tabularx}
\end{table}

\begin{figure*}[ht!]
  \centering
  \includegraphics[width=0.90\linewidth]{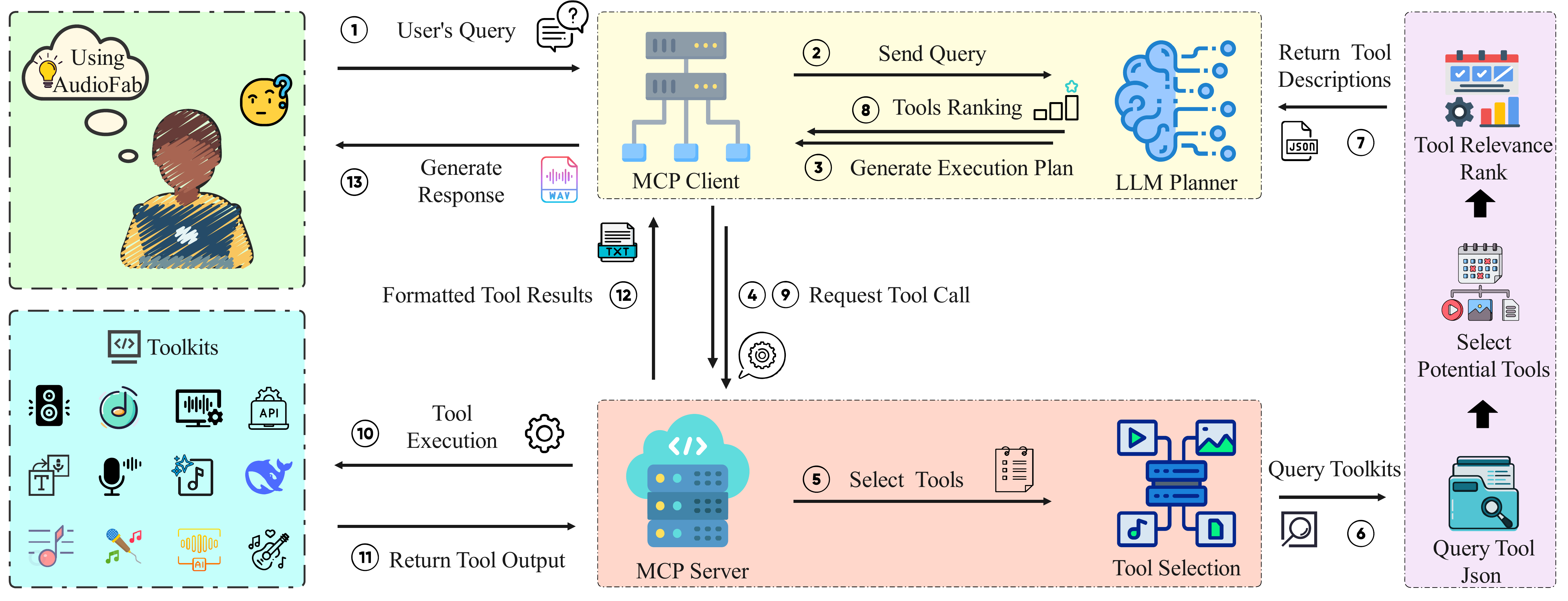}
  \caption{\textbf{Overview of the AudioFab pipeline.} Colored blocks represent architectural components, while arrows illustrate the workflow. With AudioFab, audio tools and end-users are connected through a comprehensive, user-friendly interface, enabling closer and more effective collaboration than ever before.}
  \label{fig:AudioFab}
\end{figure*}

\begin{table*}[!ht]
  \centering
    \caption{\textbf{\textit{AudioFab} supports 36 key audio techniques}. Techniques are grouped into \textit{editing}, \textit{understanding}, and \textit{generation} (further split into \textit{Audio2Audio}, \textit{X2Audio}, and \textit{Audio2X}). Capabilities are shown across \textit{speech}, \textit{general sound}, and \textit{music} domains, respectively. }

  \begin{adjustbox}{width=\textwidth,center}
    \newcolumntype{P}{>{\centering\arraybackslash}p{2cm}}
    \begin{tabularx}{\textwidth}{@{}l
        >{\centering\arraybackslash}X  
        >{\centering\arraybackslash}X  
        P P P 
      @{}}
      \toprule
      \multirow{2}{*}{\textbf{Modal}}
      & \multirow{2}{*}{\textbf{Editing}}
      & \multirow{2}{*}{\textbf{Understanding}}
      & \multicolumn{3}{c}{\textbf{Generation}} \\
      \cmidrule(lr){4-6}
      & & & \textbf{Audio2Audio} & \textbf{X2Audio} & \textbf{Audio2X} \\
      \midrule
      \textbf{Speech}
      &  Speech Editing, Speech Enhancement, Voice Activity Detection, Speech Super Resolution
      & Speech Translation, Speech Emotion Recognition, Speaker Diarization, Speaking Style Recognition
      & Speech2Song, Spoken Dialogue, Voice Conversion
      & Text2Speech, Target Speaker Extraction
      & ASR, AAC, Speech2Talking-head
      \\[0.5em]
      \midrule
      \textbf{Sound}
      & Digital Signal Processing, Audio Reconstruction, Audio Separation
      & Sound Event Detection, Audio Quality Analysis, Acoustic Scene Classification
      & Sound Style Transfer
      & Text2Audio, Video2Audio 
      & Audio2Video, Audio2Image
      \\[0.5em]
      \midrule
      \textbf{Music}
      & Music Separation, Music Mix Track, Music Format Conversion
      & Music Emotion Recognition, Music Style Description
      & Music2Song
      & Text2Music, Lyrics2Song
      & Lyrics Recognition
      \\
      \bottomrule
    \end{tabularx}
  \end{adjustbox}
 \hfill  ASR: Automatic Speech Recognition, AAC: Automated Audio Captioning
  \label{tab:1}
\end{table*}

\section{Related Work}
Several initial efforts in this area have been made by prior works. For instance, AudioGPT was first introduced in~\cite{huang2024audiogpt}. It integrates multiple specialized neural network models and complements LLM for task classification based on user queries, thus supporting understanding and generating audio modality. Likewise, frameworks such as  WavCraft \cite{liang2024wavcraft} and WavJourney \cite{liu2025wavjourney} register multiple model-based tools as functions and leverage the programming capabilities of LLM to automatically generate Python scripts for audio processing. 

While these systems demonstrate the feasibility of combining LLMs with audio tools, they face three key limitations. \textit{First}, they do not span the full lifecycle or cover the broad spectrum of audio analysis tasks. \textit{Second}, extensibility is limited—for example, WavJourney depends on structured audio scripts and script compilers, making it difficult to expand functionality or incorporate new tools. \textit{Third}, most frameworks provide only a single pre-registered model per task, restricting adaptability to emerging tasks or alternative algorithms. Overall, the absence of a modular architecture and platform support limits their flexibility and scalability.

In contrast, our work, \textbf{AudioFab}, introduces notable advancements in implementation strategy, functional scope, and module management. It supports the full workflow, from audio processing and analysis to generation and conversion, offering a comprehensive suite of capabilities. Additionally, we propose a dedicated module management mechanism to enhance extensibility and control. A detailed comparison is provided in Table~\ref{tab:2}.

\section{AudioFab: Design \& Features}
\label{sec:design}

\label{sec:audiofab-archi}
Figure~\ref{fig:AudioFab} presents the architecture of AudioFab, which comprises three core module components: the yellow block (MCP Client and LLM Planner), the orange block (MCP Server and Tool Selection), and the pink block (Process Selection). These components connect end-users (green blocks) with audio tools (blue blocks). Powered by a user-friendly, LLM-driven interface, the system enables collaborative and interoperable workflows, effectively supporting complex audio processing tasks.
Next, we provide a detailed discussion of its key features, including the module management, tool learning process, and functional scope within the audio domain.

\subsection{Module Management}
\label{sec:design1}
AudioFab adopts the recent \textit{MCP protocol}~\cite{hou2025model} to support its highly modular architecture by standardizing how audio tools interact with LLMs. In particular, the \textit{MCP Client} serves as a unified external interface, facilitating communication between users, the LLM, and the MCP Server. Meanwhile, the \textit{MCP Server} acts as the internal management hub, coordinating tool execution across various environments using standardized protocols. Each tool operates in an isolated environment (e.\,g., Conda, Docker). This prevents dependency conflicts and improves system stability and maintainability.

\subsection{Tool Learning}
\label{sec:design1}
A key innovation that sets AudioFab apart from comparable systems is its tool learning process, embedded within the full workflow. This process unfolds in four sequential stages.
Below, we describe each stage and reference the corresponding steps illustrated in Figure~\ref{fig:AudioFab}.

\textit{\textbf{Task Planning:}}
at this stage, the MCP Client serves as the user interface, receiving user commands (\textit{step 1}) and forwarding them to the LLM (\textit{step 2}). The LLM, as the system’s core planner, interprets the request, decomposes it into structured subtasks, and returns them to the client (\textit{step 3}). Unlike traditional linear ReACT processes, AudioFab employs specialized prompts and a feedback mechanism tailored for audio tasks, enabling dynamic evaluation and adjustment for improved robustness and accuracy.

\textit{\textbf{Tool Selection:}}
next, the MCP Client forwards the request to the MCP Server (\textit{step 4}), which acts as an adapter by translating tool calls into commands the tools can understand. These commands are then passed to the Tool Selection Module (\textit{step 5}). This module includes three parts: tool enumeration, retrieval matching, and parameter querying. It uses a JSON file containing tool names, descriptions, and examples. During initialization, only tool instructions are loaded into the LLM to save tokens. When a user issues a tool-related request (\textit{step 6}), semantic matching identifies the most relevant toolchains, returning corresponding usage instructions and examples (\textit{step 7}). With few-shot learning, this helps ensure accurate selection and reduces errors or hallucinations.

\textit{\textbf{Tool Invocation:}} this stage includes two phases, namely Tool Call and Tool Execution. The LLM evaluates the toolchain provided by the selection module and submits a structured request (\textit{step 8}). The MCP Server handles the request (\textit{step 9}) through unified invocation and execution (\textit{step 10}), then returns the tool outputs back to the server (\textit{step 11}).

\textit{\textbf{Response Generation:}}
finally, once all subtasks are completed, formatted tool results are sent to the LLM via the MCP Client (\textit{step 12}). The LLM aggregates and synthesizes these outputs into a coherent response and returns it to the user through the MCP Client (\textit{step 13}), completing the end-to-end workflow.

\subsection{Comprehensive and Growing Capabilities}

Building on the modular architecture and tool learning mechanisms described above, AudioFab enables flexible integration, expansion, and management of diverse audio toolkits. The system currently supports 36 audio tasks, spanning from basic operations like format conversion and noise reduction to advanced applications such as voice conversion and cross-modal generation. These capabilities, summarized in Table~\ref{tab:1}, reflect the system’s breadth and depth. Furthermore, AudioFab is designed to continuously evolve, allowing seamless addition of new tools as needs arise. \textit{We believe this is just the beginning of AudioFab’s journey, with ample potential for future expansion and continued impact in the audio AI and multimodal research community.}

\setlength{\parskip}{0pt}

\section{{Evaluation with AudioFab}}
\label{sec:exp}

\begin{figure*}[!t]
    \includegraphics[width=0.997\linewidth, trim=0 105 0 0, clip]{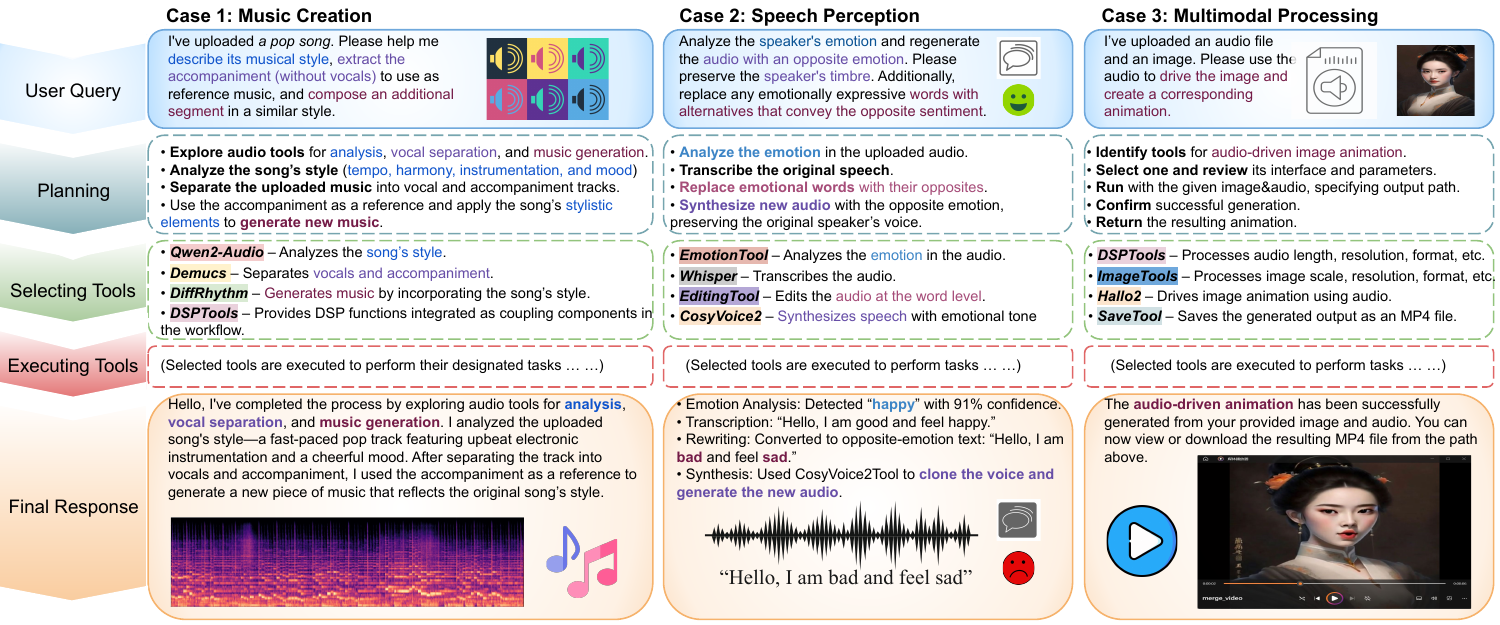}
      \caption{\textbf{Three illustrative usage scenarios of the AudioFab toolkit: (1) music creation, (2) speech editing, and (3) multimodal interaction.} Each scenario demonstrates a prototypical workflow—from user query and plan generation to tool selection and final output, showcasing the toolkit’s versatility, structured execution, and practical applicability.}
      \label{fig:examples_all_v1}
\end{figure*}

Due to the absence of standardized benchmarks for evaluating AudioFab’s broad capabilities, we assess its performance through three illustrative examples, each representing a key use case.
These diverse scenarios are carefully chosen to: (i) test across different modalities, (ii) validate its support for chained workflows involving multiple interconnected tools, and (iii) show its ability to edit, understand, and generate audio. The versatility of AudioFab is demonstrated through these examples, as illustrated in Figure~\ref{fig:examples_all_v1}.

\textit{\textbf{Music Creation:}} this case showcases AudioFab's music creation capabilities. Given a pop song, the system analyzes its style, separates vocals from accompaniment, and generates new segments in a similar style using the accompaniment as reference.

\textit{\textbf{Speech Perception:}} this case indicates AudioFab's speech perception capabilities. Given a speech clip, the system recognizes emotion, transcribes the content, replaces emotional terms with their opposites, and regenerates speech with reversed emotion while preserving the original timbre. 

\textit{\textbf{Multimodal Processing:}} this case highlights AudioFab’s multimodal processing capabilities. Given an image and audio input, the system generates audio-driven animation by transforming the image based on audio dynamics and synthesizing a coherent video.

In summary, these three use cases confirm AudioFab’s effectiveness as a specialized toolset and tool learning solution, showcasing its potential as a unified platform for intelligent audio processing.

\section{Conclusion and Future Work}
\label{sec:conclusion}



In this paper, we present \textit{AudioFab}, an open-source unified audio agent framework covering three core modalities: speech, sound, and music. It integrates 36 key functionalities through an intuitive natural language interface. 
The tool serves two key audiences: for \textit{non-experts}, it eliminates the need for specialized domain knowledge to handle complex audio tasks, while for \textit{researchers and developers}, it reduces overhead in tool deployment, facilitating deeper exploration and more rapid prototyping. Ultimately, this solution lowers the field's entry barrier for practitioners across all levels, making audio technologies accessible to a broader audience.

In the future, we aim to establish a comprehensive benchmark and leaderboard to assess the framework's reliability and quality, maintained and updated on GitHub.

\begin{acks}

This work was supported by the Guangdong Basic and Applied Basic Research Foundation under Grant No. 2024A1515010112, the Changsha Science and Technology Bureau Foundation under Grant No. kq2402082, and the National Natural Science Foundation of China under Grant No. 62076092.
\end{acks}

\bibliographystyle{ACM-Reference-Format}
\bibliography{sample-base}


\end{document}